\documentstyle[12pt,aasms4]{article}

\begin{document}

\title{The Importance of Nonlinear H$_{2}$ Photoexcitation in Strongly
Irradiated PDRs}

\author{P.P Sorokin and J.H. Glownia}
\setlength{\baselineskip}{0.5\normalbaselineskip}
\affil{IBM Research Division \\P.O. Box 218, Yorktown Heights, NY 10598-0218\\ sorokin@us.ibm.com; glownia@us.ibm.com}
\authoraddr{ P.O. Box 218, Yorktown Heights, NY 10598-0218}
\authoremail{sorokin@us.ibm.com; glownia@us.ibm.com}
\and
\author{R.T. Hodgson} 
\affil{822 Pinesbridge Rd., Ossining, NY 10562\\patents@aip.org}
\authoraddr{822 Pinesbridge Rd., Ossining, NY 10562}
\authoremail{patents@aip.org}

\begin{abstract}
It is shown that, under sufficiently intense OB-star illumination of a
stationary photoexcitation front (PDR), nonlinear H$_{2}$ photoexcitation
processes comprising driven resonant two-photon transitions between X-state
quantum levels, with VUV continuum light from the star supplying both
driving fields, largely determine the photonic pathways of H$_{2}$ molecules
in the PDR close to the ionization front. Specifically, for a flux of $%
\sim $4 x 10$^{5}$ Habing fields incident upon a PDR, the total rate at
which an H$_{2}$ molecule is nonlinearly photoexcited out of any X-state
quantum level is calculated to be roughly 100 times greater than the total
rate at which it is linearly photoexcited out of the same level. In strongly
excited PDRs, the populations in almost all of the $\sim$300 bound
quantum levels of the X state will be maintained approximately equal via a
few myriads of interconnecting two-photon steps. The remarkable importance
of two-photon transitions in H$_{2}$ photoexcitation in strongly irradiated
PDRs derives from the exceptionally narrow Raman linewidth ($\Gamma
\sim $10$^{-6}$ $\sec ^{-1}$) that characterizes all two-photon
transitions between bound H$_{2}$ X-state quantum levels.

\end{abstract}

\keywords{infrared: ISM - - - ISM: clouds - - - ISM: HII regions - - - ISM: lines - - - ISM: lines and bands - - - ultraviolet: ISM}

\section{Introduction}
In existing theoretical models of stationary photoexcitation fronts (PDRs)
\setlength{\baselineskip}{0.5\normalbaselineskip}
$[$see Draine \& Bertoldi (1996) for a comprehensive list of references$]$, a
one-dimensional geometry is usually considered, with light from an OB-type
illuminating star assumed incident upon the PDR from one direction. In all
published PDR models, only linear H$_{2}$ photoexcitation processes have
been considered. The consensus among astrophysicists is that these models
adequately describe the main photonic events that occur in H$_{2}$%
-containing clouds separated by at least 10 pc from OB-type illuminating
stars. In the present Letter, we draw attention to the fact that, in very
strongly irradiated PDRs (for example, those with cloud-to-star distances $%
\leq $0.1 pc), nonlinear photoexcitation of H$_{2}$ can evidently become the
dominant photonic process determining the structure of the PDR.

\section{Resonantly Enhanced, VUV Starlight-Driven, Two-Photon Transitions between
H$_{2}$ X-State Quantum Levels}

An estimate of the resonantly-enhanced, VUV starlight-driven, two-photon
transition rate between two H$_{2}$ X-state quantum levels can be obtained
with use of the formula shown in Fig. 1. In SI units, this formula $[$c.f.
eq. (5.6b) of Hanna et al. (1979)$]$ gives the cross-section for Stokes-wave
gain in resonant stimulated Raman scattering (SRS) in a three-level system.
However, it equally well gives the cross-section for induced nonlinear
absorption at $\omega _{P}$ when an intense radiation field is present at $%
\omega _{S}$ - a process often referred to as inverse Raman scattering
(IRS). It has long been recognized (Jone \& Stoicheff 1964; MacQuillan \& Stoicheff 1966) that both processes are fundamentally the same. In this
equation, the classical electron radius $r_{e}$ $\approx $ 2.82 x 10$%
^{-15}$ m. The quantity $I_{P}$ is the pump intensity in W/m$^{2}$. The
subscripts $g,i,$ and $f$ designate the ground level, intermediate level,
and final level, respectively, of the three-level system. The quantity $%
\Gamma $ is the Raman linewidth (in units of angular frequency), which, for
the astrophysical environment assumed in our PDR model, would be entirely
determined by a convolution of the vibrational decay rates of the two Xn$^{\prime\prime}$
quantum levels involved. In light of recent calculations of quadrupole
transition probabilities for all X-state radiative decays (Wolniewicz et al. 1998), one can realistically assume a value $\Gamma $
= 10$^{-6}$ $\sec ^{-1}$ for all two-photon transitions between X-state
quantum levels. The quantities $f_{fi}$ and $f_{gi}$ are the oscillator
strengths of the Stokes-wave and primary frequency transitions. For all H$%
_{2}$ Lyman (B$\Longleftrightarrow $X) and Werner (C$\Longleftrightarrow $X)
transitions, these $f$- values can be simply determined from the tables in
Abgrall et al. (1993 a,b). The quantities such as $\Omega _{if}$ are the
energy separations of the levels involved, expressed as angular frequencies,
e.g. $(h/2\pi )\Omega _{if}$ = $E_{i}$ - $E_{f}$. Note that the Stokes-wave
frequency $\omega _{S}$ essentially cancels $\Omega _{if}$ in the present
case. The equation in Fig. 1 gives the cross-section in units of m$^{2}$.

To estimate the rates of both nonlinear and linear H$_{2}$ photoexcitation
in the PDR, one needs to know the VUV flux from the star incident upon the
cloud surface. Here, for definiteness, we assume the illuminating star to be
a B0 III star, with temperature T=31,500K and radius R = 16$\Re _{\odot }$.
We take the cloud-to-star distance to be 0.1 pc, and assume that there is no
intervening dust cloud between the star and H$_{2}$-containing cloud. From
the Planck formula for emittance of photons from a blackbody per cm$^{2}$
per sec per unit frequency range (Allen 1976), one finds that at 1,000 \AA\ %
the incident flux on the PDR is $\phi _{1000}$ = 2.55 x 10$^{8}$ photons per
cm$^{2}$ per sec per cm$^{-1}$. Habing (1968) estimated the intensity of
interstellar starlight at $\lambda $=1,000 \AA\ to be $\lambda u_{\lambda }$
= 4 x 10$^{-14}$ erg cm$^{-3}$. At the surface of the PDR we are here
considering, therefore, $\phi _{1000}$ is about 4.25 x 10$^{5}$ Habing
fields.

As is the case in almost all published PDR models, it is here assumed that
ionizing radiation (i.e. $\lambda $ $\leq $ 912 \AA ) from the star
creates a thin H II region on the surface of the PDR cloud nearest the star.
Within this H II region, a large fraction ($\sim $2/3) of the incident
ionizing photons are converted to Ly-$\alpha $ photons via H$^{+}$+ e$^{-}$
recombinations followed by cascading transitions through H-atom excited
states. Half of the Ly-$\alpha $ photons enter the neutral region, with
their frequencies having become distributed in a $\sim $20-cm$^{-1}$%
-wide spectral band via the frequency redistribution that results from
elastic scattering by hot H atoms in the H II region. One can easily
estimate the total flux of ionizing photons incident upon the PDR from
tables of the Planck function(Allen 1976). One finds this to be $\phi
_{ion}$ $\approx $ 6.2 x 10$^{12}$ photons cm$^{-2}$ $\sec ^{-1}$.
There are thus about $\phi _{Ly\alpha }$ $\approx $ 2.1 x 10$^{12}$ Ly-$%
\alpha $ photons entering the neutral PDR region per cm$^{2}$ per second.
Thus, the Ly-$\alpha $ flux (per unit wavenumber) entering the neutral PDR
region is seen to exceed that at 1,000 \AA\ by roughly a factor 400. In $\S $%
3, we will briefly hint at an important effect this Ly-$\alpha $ radiation
should have on the nonlinear excitation of H$_{2}$ in strongly irradiated
PDRs.

To estimate the total rate at which an H$_{2}$ molecule in the PDR would
undergo linear photoexcitation, one can use Table 2 of Draine \& Bertoldi (1996). These authors have calculated and summed individual contributions to
the total unshielded H$_{2}$ linear photoexcitation rate from all X$%
\rightarrow $B and X$\rightarrow $C transitions. For a flux $\phi _{1000}$ =
4.25 x 10$^{5}$ Habing fields, the total linear photoexcitation rate for a
molecule in (X0, J$^{\prime\prime}$=1) would be about 1.2 x 10$^{-4}$ $\sec ^{-1}$. In the
PDR model of Draine \& Bertoldi (1996), a Doppler width $\approx $1.7 cm%
$^{-1}$ is assumed.

With use of the equation shown in Fig. 1, we now evaluate the driven
(simultaneous, not sequential) two-photon transition rate from X0,
J$^{\prime\prime}$=1 to X1, J$^{\prime\prime}$=1 
involving C0($\Pi _{u}^{-}$), J$^{\prime}$=1 as resonant
intermediate state. A rough - but conservative - approximation one can make
in integrating the expression for $\sigma _{IRS}$ over the applied VUV
pumping continuum is to assign to the quantity ($\Omega _{ig}$ - $\omega _{P}
$) appearing in the denominator of the equation in Fig. 1 a constant value
corresponding to an offset of one wavenumber, and then integrate over a
spectral interval two wavenumbers wide on either side of the resonance. From
the value of $\phi _{1000}$ earlier estimated, the value of $I_{P}$ for a
four-wavenumbers-wide spectral interval is found to be 0.00002 W/m$^{2}$.
With substitution of appropriate values for the other quantitities appearing
in the equation of Fig. 1 (e.g. $f_{gi}$ = 0.03, $f_{fi}$ = 0.07), one finds
the cross-section for induced absorption to be $\sigma _{IRS}$ $\approx 
$ 8.4 x 10$^{-14}$ cm$^{2}$. Multiplying this value by $\phi _{1000}$ x 4
shows the transition rate of the IRS process to be $\sim $8.6 x 10$^{-5}
$ $\sec ^{-1}$. The rate of nonlinear photoexcitation of an H$_{2}$ molecule
in (X0, J$^{\prime\prime}$=1) via this particular two-photon transition is thus about 0.7
times the earlier estimated total rate of linear photoexcitation of an H$_{2}
$ molecule in the same (X0, J$^{\prime\prime}$=1) quantum level.

However, driven two-photon transitions from (X0, J$^{\prime\prime}$=1) to 
(X1, J$^{\prime\prime}$=1) can
also occur via the paired transitions Cn-0Q1, Cn-1Q1 (n = 1-5) and also
Cn-0R1, Cn-1R1 (n = 0-5). This makes the total two-photon transition rate
out of (X0, J$^{\prime\prime}$=1) exceed the total linear rate by $\sim $7 times. (The
wavelengths of both C6-0Q1 and C6-0R1 are shorter than 912 \AA , the Lyman
limit. No $\lambda $ $\leq $ 912 \AA\ starlight can penetrate beyond
the HII region of the PDR.) One can also go from 
(X0, J$^{\prime\prime}$=1) to (X1, J$^{\prime\prime}$=3)
via transitions Cn-0R1, Cn-1P3. With the inclusion of these paths, the
two-photon transition rate out of (X0, J$^{\prime\prime}$=1) would be roughly 10 times the
total linear rate of photoexcitation. One next must consider analagous
two-photon transitions from (X0, J$^{\prime\prime}$=1) that terminate on
 (X2, J$^{\prime\prime}$=1), 
(X2, J$^{\prime\prime}$=3), ...(X8, J$^{\prime\prime}$=1), (X8, J$^{\prime\prime}$=3), etc..
With inclusion of these, the
two-photon excitation rate from (X0, J$^{\prime\prime}$=1) would exceed the total linear
excitation rate by $\sim $70 times. Finally, two-photon excitation from
(X0, J$^{\prime\prime}$=1) can also occur via B-state resonant intermediate levels.
Additional inclusion of such pathways, makes the total rate of two-photon
excitation out of (X0, J$^{\prime\prime}$=1) exceed the total linear photoexcitation rate
out of the same quantum level by at least a factor 100. Roughly the same
nonlinear-to-linear photoexcitation rate ratio would obtain for H$_{2}$
molecules in any of the $\approx $300 X-state quantum levels. The
estimated total nonlinear photoexcitation rate out of any such level is thus 
$\sim $1.2 x 10$^{-2}$ $\sec ^{-1}$.

\section{Consequences of VUV Starlight-Driven Two-Photon Transitions Occurring in Heavily
Irradiated PDRs}

From the transition rates calculated above, it is apparent that, through a
few myriads of interconnecting two-photon transitions, the H$_{2}$
populations in each of the $\approx $300 X-state quantum levels would be
maintained approximately equal. Note that no H$_{2}$ photodissociation can
result from two-photon transitions occurring from the highest bound X-state
levels to unbound X-state continuum levels. The two-photon transition rate
depends inversely upon the Raman linewidth, and the latter represents a
convolution of the lifetimes of the quantum levels connected by the
two-photon process. When one steps from the highest bound X-state levels to
unbound levels of the X-state continuum, the lifetime drops by roughly
eighteen orders of magnitude! On the basis of this same disparity in
lifetimes, one perhaps should question the validity of the
assumption usually made in linear PDR models that roughly one out of ten H$%
_{2}$ photoexcitations from X-state quantum levels result in
photodissociation via fluorescent VUV transitions terminating on unbound
levels of the X-state. This would be true if linear photoexcitation of H$_{2}
$ in PDRs did involve real excitation of B- and C-state quantum levels.
However, in collisionless media, it is known (\cite{LOUDON}) that linear
photoexcitation occurs entirely via spontaneous resonant Raman scattering,
and, at exact resonance, the cross-section for the latter varies again as $%
\Gamma ^{-1}$ (Shen 1974).

If one knows the total H$_{2}$ density in the neutral region of the PDR, one
can approximately estimate the maximum thickness $d$ of the nonlinearly
excited region as follows. Assume an H$_{2}$ density of 10$^{4}~$cm$^{-3}$.  
In the nonlinearly excited region, these molecules will be evenly
distributed amongst all the $\approx $300 X-state quantum levels, with
the spontaneous infrared radiation emission rate from each level being $%
\sim $10$^{-6}$ $\sec ^{-1}$. Assume the average IR emitted photon to
have an energy of 5,000 cm$^{-1}$. Equating the VUV power from the star
incident upon the PDR in the entire spectral range of the Lyman and Werner
bands that lies longward of 912 \AA\ to the IR power radiated in a column of
thickness $d$:
\begin{displaymath}
(20,000)(2.55\  \mbox{x} \  10^{8})(100,000)\approx (10^{4})(10^{-6})(5,000)d,
\end{displaymath}
one finds d$\approx $0.003 pc. Figure 1 of \cite{LEMAIRE} is a
striking colored image (colors corresponding to emission intensity) of H$_{2}
$ 1-0 S(1) vibrational emission at 2.121 $\mu $m in NGC 7023, taken at very
high spatial resolution. It shows H$_{2}$ vibrational emission originating
from a $\sim $0.02-pc-thick, roughly planar, cloud located $\sim $%
0.07 pc to the NW of the illuminating star HD 200775. (From Earth, one views
this cloud largely through one of its `edges'.) There is very much higher H$%
_{2}$ emission coming from three $\sim $0.004-pc-thick embedded
filaments located near the surface of the cloud facing the star. In view of
the calculations that have been presented above, we suggest that these
bright filaments in NGC 7023 might represent PDR regions that are
nonlinearly excited.

It was noted above that Ly-$\alpha $ radiation generated in the H II region
enters the PDR neutral region at an intensity level roughly 400 times
greater than the continuum level at 1,000 \AA, but only over an estimated $%
\sim $20-cm$^{-1}$-wide bandwidth. There are a dozen or so allowed
transitions originating from X-state quantum levels whose frequencies fall
within such a Ly-$\alpha $ spectral distribution. A slight contribution
towards the equalization of H$_{2}$ populations in the various X-state
quantum levels arises from these resonances. However, there is a different -
rather profound - effect that this Ly-$\alpha $ radiation can induce, when
it is applied to an equally populated manifold of X-state quantum levels. It
can induce broadband stimulated Raman scattering (SRS) to occur on the three
or four strongest of the above mentioned resonant transitions, generating
broadband IR Stokes-wave light on strong transitions to EF-states. An SRS
process would occur as part of a 2n-wave parametric oscillation (SRS-PO)
process, with additional IR and VUV light being generated on strong
transitions ultimately returning molecules to the X-state levels from which
the SRS-PO processes originated. As shown elsewhere (Sorokin \& Glownia 2001), the so-called `unidentified infrared (emission) bands (UIBs)' can be
successfully assigned to IR emissions coherently generated in such SRS-PO
processes.

\acknowledgments
One of the authors (PPS) acknowledges receiving strong encouragement from
Dr. Anita J. Schell-Sorokin to publish the central idea upon which this
article is based.

\clearpage
\begin{figure}
\plotone{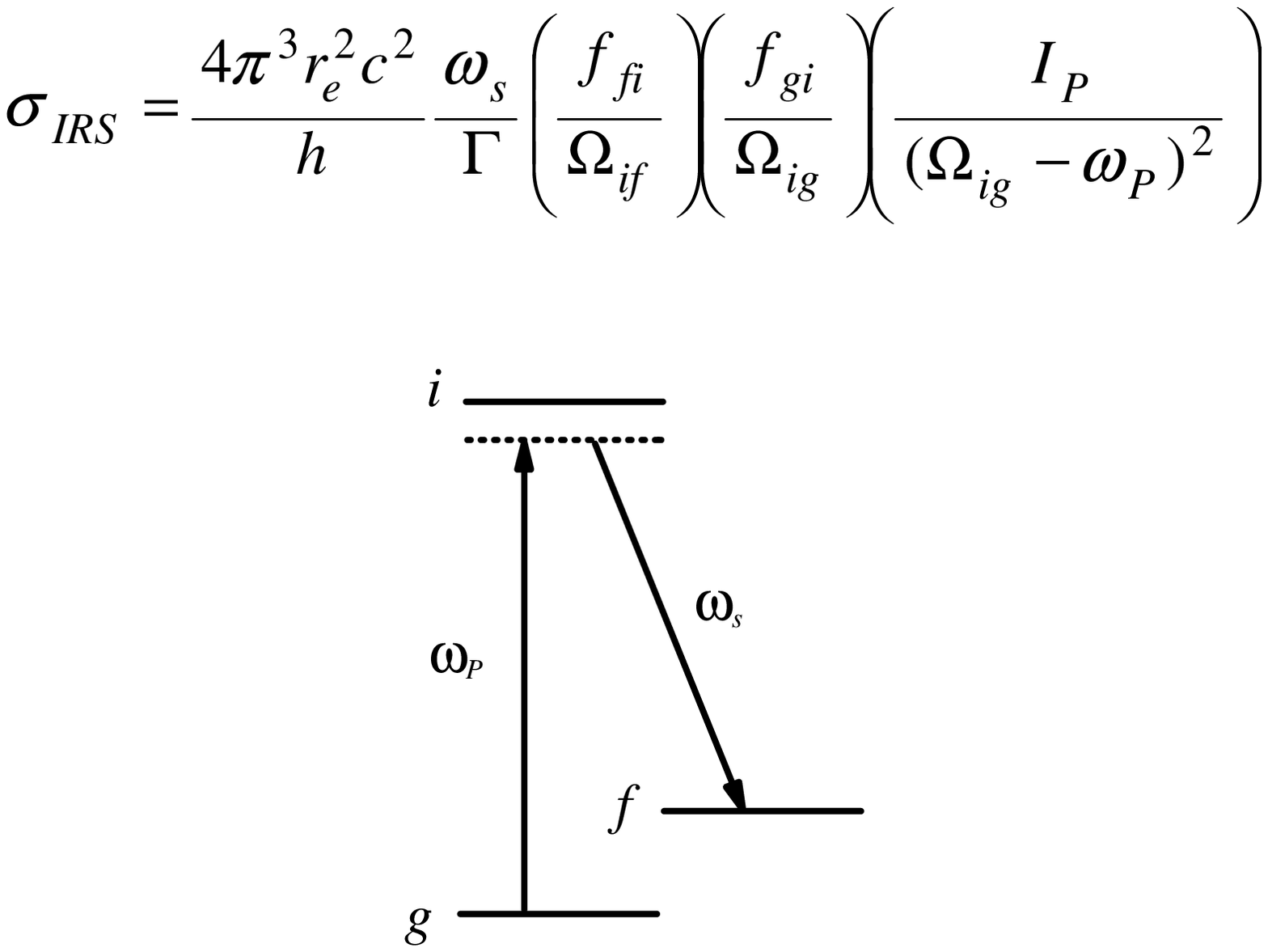}
\caption{Equation for the cross-section of either SRS Stokes-wave gain or IRS-induced absorption in a resonant, three-level system.  The light waves in the diagram are depicted for the SRS case.}
\end{figure}

\begin{thebibliography}{}
\bibitem[Abgrall et al. (1993a)]{ABGRALL1}Abgrall, H., Roueff, E., Launay, F., Roncin, J.-Y., \& Subtil, J.-L.
1993a,  A\&AS, 101, 273.
\bibitem[Abgrall et al. (1993b)]{ABGRALL2}--------.
1993b, A\&AS, 101, 323.
\bibitem[Allen 1976]{ALLEN}Allen, C.W. 1976, Astrophysical Quantities (London and Dover, New Hampshire: The Athlone Press).
\bibitem[Draine \& Bertoldi (1996)]{DRAINE}Draine, B.T. \& Bertoldi, F. 1996, \apj, 468, 269.
\bibitem[Habing (1968)]{HABING}Habing, H. J. 1968, \bain, 19, 421.
\bibitem[Hanna et al. (1979)]{HANNA}Hanna, D.C., Yuratich, M.A., \& Cotter, D. 1979, Nonlinear Optics of Free Atoms and Molecules (Berlin, Heidelberg, \& New York: Springer-Verlag).
\bibitem[Jones \& Stoicheff 1964]{JONES}Jones, W.J. \& Stoicheff, B.P. 1964, \prl, 13, 657.
\bibitem[Lemaire et al. (1996)]{LEMAIRE}Lemaire, J.L., Field, D., Gerin, M., Leach, S., Pineau des For\^{e}ts, G., Rostas, F., \& Rouan, D. 1996, \aap, 308, 895.
\bibitem[Loudon 1983]{LOUDON}Loudon, R. 1983, The Quantum Theory of Light (2d ed., Oxford: Clarendon Press).
\bibitem[MacQuillan \& Stoicheff 1966]{MACQUILLAN}MacQuillan, A.K. \& Stoicheff, B.P. 1966, in Physics of Quantum Electronics, ed. P.L. Kelley, B. Lax, \& P.E. Tannenwald (New York: McGraw-Hill), 192.
\bibitem[Shen (1974)]{SHEN}Shen, Y.R. 1974, \prb, 9, 622.
\bibitem[Sorokin \& Glownia 2001]{SOROKIN}Sorokin, P.P. \& Glownia, J.H. 2001, to be published.
\bibitem[Wolniewicz et al. 1998]{WOLN}Wolniewicz, L., Simbotin, I., \& Dalgarno, A. 1998, \apjs, 115, 293.
\end{thebibliography}
\end{document}